\documentclass[prb,aps,twocolumn,preprintnumbers,amsmath,amssymb,superscriptaddress,showpacs]{revtex4-2}

\usepackage[colorlinks,bookmarks=true,citecolor=blue,linkcolor=red,urlcolor=blue,citecolor=blue]{hyperref}

\usepackage{graphicx}
\usepackage{epstopdf}
\usepackage{amsmath}        % 数学符号
\usepackage{multirow}
\usepackage{xcolor}
\usepackage{ulem}
\usepackage[version=4]{mhchem}
\usepackage{enumitem}
\usepackage{booktabs}  % 三线表核心宏包
\usepackage{tikz}
\usepackage{siunitx}       % 数值对齐（可选）
\usepackage{times}         % Times New Roman字体
      
\newcommand{\RNum}[1]{\uppercase\expandafter{\romannumeral #1\relax}}

\begin{document}
%\linenumbers

\title {Tree tensor network impurity solver based on Cayley-tree mapping}

\author{Bo Zhan}
\author{Jia-Lin Chen}
\author{Zhen Fan}
\author{Tao Xiang}
\email{txiang@iphy.ac.cn}
\address{Beijing National Laboratory for Condensed Matter Physics and Institute of Physics,
Chinese Academy of Sciences, Beijing 100190, China.}
\address{School of Physical Sciences, University of Chinese Academy of Sciences, Beijing 100049, China}
\date{\today}

\begin{abstract}
  
  We introduce a tree tensor network (TTN) impurity solver that enables highly efficient and accurate real-time simulations of quantum impurity models. By decomposing a noninteracting bath Hamiltonian into a Cayley tree, the method provides a tensor network representation that naturally captures the multiscale entanglement structure intrinsic to impurity–bath systems. This geometry differs from conventional chain-based mappings and yields a substantial reduction of entanglement, allowing accurate ground-state properties and long-time dynamics to be captured at significantly lower bond dimensions. Benchmark calculations for the single-impurity Anderson model demonstrate that the TTN solver achieves markedly enhanced resolution of real-frequency spectral functions, without  invoking analytic continuation. This impurity solver provides a balanced, scale-uniform description of impurity physics and offers a versatile approach for real-time dynamical mean-field theory and related applications involving quantum impurity models.
  
\end{abstract}
\maketitle

\section{Introduction}

  Quantum impurity models constitute a central class of paradigmatic problems in condensed matter physics~\cite{wilson1975,Hewson_1993,coleman2007}. By describing localized degrees of freedom coupled to extended environments, they capture the essential physics underlying a wide range of strongly correlated phenomena. Historically, the study of magnetic impurities embedded in metallic hosts, most prominently exemplified by the Kondo problem, revealed how the interplay between localized moments and itinerant electrons gives rise to highly nontrivial many-body behavior, including the formation of an emergent low-energy scale characterized by the Kondo temperature and nonperturbative screening effects~\cite{Kondo1964,Anderson_1970,Anderson1961}. Since then, quantum impurity models have evolved into indispensable theoretical laboratories, ranging from the single-impurity Anderson model~\cite{Anderson1961,Wolff1961,Schrieffer1966} to multi-impurity and multi-orbital generalizations relevant to heavy fermion systems, nanostructures, and correlated materials~\cite{Coqblin1969,Nozieres1980KondoEI,Stewart1984,Goldhaber-Gordon1998}. Accurately solving these models, therefore, remains a challenge of fundamental and practical importance.

 The significance of quantum impurity models has been further elevated by the development of dynamical mean-field theory (DMFT)~\cite{Metzner1989,Georges1992,Georges1996}. DMFT maps a correlated lattice model in the limit of large coordination number onto a self-consistent quantum impurity problem embedded in a dynamical bath, thereby capturing local quantum fluctuations exactly while treating spatial correlations at the mean-field level. This nonperturbative framework has proven remarkably successful across a broad range of interaction strengths and temperatures. Moreover, when combined with density functional theory~\cite{Hohenberg1964,Jones1989}, DMFT has become a powerful {\it ab initio} approach for studying the electronic structure of real correlated materials~\cite{Anisimov_1997, Lichtenstein1998, Kotliar2006,Held2007}. In all practical implementations of DMFT, however, the accuracy and efficiency of the calculation hinge critically on the availability of a reliable quantum impurity solver~\cite{Georges1996}. Since DMFT self-consistency typically requires many iterations, the impurity solver must simultaneously deliver high precision and manageable computational cost, making solver development a long-standing and actively pursued challenge.

Over the years, a variety of impurity solvers have been developed, each with distinct strengths and limitations. Continuous-time quantum Monte Carlo (CTQMC)~\cite{Rubtsov2005,Werner2006,Gull2011} methods are numerically exact on the imaginary-time or Matsubara-frequency axis and can efficiently treat multi-orbital interactions. However, their applicability to real-frequency dynamics is limited by the fermionic sign problem and by the ill-posed nature of analytic continuation. Exact diagonalization (ED)~\cite{Caffarel1994,Capone2007,Granath2012,Lu2014} provides direct access to real-frequency spectra but is severely restricted by the size of its accessible Hilbert space. The numerical renormalization group (NRG)~\cite{wilson1975,Bulla1999,Bulla2001,Bulla2008} excels in resolving low-energy properties and satisfies the Friedel sum rule, yet it offers limited resolution at higher energies~\cite{Peters2006}. The natural orbital renormalization group (NORG)~\cite{He2014} reduces the effective Hilbert space via truncation in a natural-orbital basis, but its performance depends on how well that basis captures the entanglement structure.

 Tensor network approaches, most notably those based on matrix product states (MPS) and density-matrix renormalization group (DMRG) techniques~\cite{White1992, Shibata1996,schollwock2005,Garcia2004,Hallberg_2006,Heidrich2009,Peters2011,Wolf2014,Wolf2015,Ganahl2015,Bauernfeind2017}, have emerged as powerful, sign-problem-free alternatives capable of accessing both imaginary- and real-time dynamics. These methods provide direct access to real-frequency quantities without analytic continuation and have been successfully applied to quantum impurity problems. However, the strictly one-dimensional structure underlying MPS representations often leads to a rapid growth of entanglement during real-time evolution, particularly when long-range bath correlations or complex connectivity are involved. As a consequence, achieving accurate long-time dynamics frequently requires prohibitively large bond dimensions, limiting computational efficiency.

 In this work, we introduce a TTN–based impurity solver that overcomes key limitations of existing approaches in real-time and real-frequency calculations. We develop a general scheme to decompose the single-impurity model onto a hierarchical Cayley-tree lattice, enabling a natural and efficient formulation of impurity problems within the TTN geometry. This representation offers greater flexibility for capturing the entanglement structure of impurity–bath systems. In particular, the tree geometry naturally accommodates the inhomogeneous and multiscale entanglement induced by the impurity coupling. As a result, our method achieves substantially higher efficiency in describing both ground-state properties and real-time dynamics, reaching comparable or improved accuracy at significantly lower bond dimensions.

 The remainder of this paper is organized as follows. In Section~\ref{Sec:Model}, we introduce the single-impurity model studied in this work and present a novel scheme that decomposes the bath Hamiltonian onto a hierarchical Cayley-tree lattice, illustrated using a binary-tree construction. In Section~\ref{Sec:results}, we present results for the ground-state properties and dynamical spectral functions obtained with the tree tensor network renormalization group method, together with a systematic comparison to MPS calculations. Finally, Section~\ref{Sec:summary} summarizes our main findings and conclusions.

\section{Model and method}
\label{Sec:Model}

 This work focuses on the single-impurity Anderson model (SIAM), which serves as a paradigmatic quantum impurity problem, although the framework developed here is readily generalizable to multi-impurity systems. The Hamiltonian of the SIAM reads
\begin{eqnarray}  \label{Eq:Ham0}
 H & = & H_b+ H_c + H_\text{imp} ,\\
 H_b &=& \sum_{k}\varepsilon _{k}c_{k}^{\dagger }c_{k} ,\\
 H_c & = &\sum_k\left( V_k d^\dagger c_k +
 V^*_k c^\dagger_k d\right), \\
 H_\text{imp} & = & \varepsilon_d n_d + U n_{d\uparrow}
n_{d\downarrow} , 
\end{eqnarray}
 The bath Hamiltonian \(H_b\) describes a set of noninteracting fermionic modes, while \(H_c\) accounts for their hybridization with the impurity. The bath consists of $N$ fermionic orbitals labeled by \(k=1,\cdots, N\), with creation (annihilation) operators \(c_k^\dagger\) (\(c_k\)) and single-particle energies \(\varepsilon_k\). The corresponding hybridization amplitudes are denoted by \(V_k\).

\begin{figure}[t]
    \centering
    \includegraphics[]{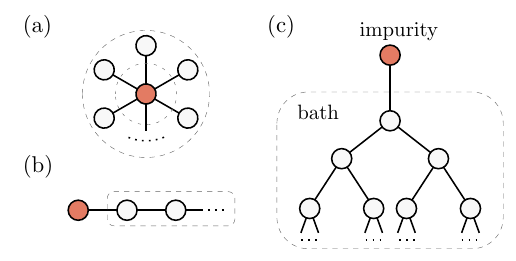}
    \caption{Three equivalent representations of the SIAM: (a) a star geometry representation, (b) a one-dimensional chain representation, and (c) a binary-tree representation. The interacting impurity site is shown in orange, while the noninteracting fermionic bath sites are shown in white. Links indicate hopping (hybridization) between sites.}
    \label{fig:bath_geometry}
\end{figure}

 As written, this Hamiltonian admits a natural representation in a “star” geometry [Fig.~\ref{fig:bath_geometry}(a)], in which all bath orbitals are directly coupled to the impurity. However, an arbitrary unitary transformation among the bath degrees of freedom produces a mathematically equivalent Hamiltonian with a different connectivity. A widely used choice is the linear “chain” geometry [Fig.~\ref{fig:bath_geometry}(b)], obtained via a Lanczos tridiagonalization~\cite{Allerdt2019}. This geometry is commonly employed in NRG and DMRG calculations. Despite their formal equivalence, different geometrical representations exhibit markedly distinct entanglement structures, which play a decisive role in determining the efficiency and accuracy of tensor-network-based simulations.

  In this work, we propose a systematic method to map the bath Hamiltonian onto a Cayley-tree [Fig.~\ref{fig:bath_geometry}(c)] and solve the SIAM using the TTN method. A Cayley tree  is a loop-free lattice in which each site is connected to a fixed number of neighbors, called the coordination number. This tree topology provides a flexible organization of the bath degrees of freedom and offers two principal advantages in solving SIAM. First, the branching structure distributes entanglement across multiple paths, thereby reducing the maximum entanglement entropy per bond and enabling a more accurate representation of the quantum state at a fixed bond dimension. Second, the hierarchical nature of the tree naturally captures the multi-scale entanglement between the impurity and a continuum bath, making it particularly well suited for describing long-range correlations and long-time dynamics, which are essential for obtaining high-resolution spectral functions.

\subsection{Cayley-tree mapping of the bath Hamiltonian}
  \label{Sec:tree_transform}

 We now introduce a systematic approach to map the bath Hamiltonian \(H_b\) onto a Cayley-tree lattice. For clarity, we illustrate the construction using a binary-branching Cayley tree (coordination number \(z=3\)), although the procedure generalizes straightforwardly to arbitrary branching numbers and tree topologies. The mapping relies only on the quadratic (noninteracting) structure of the bath Hamiltonian and therefore applies to any noninteracting fermionic bath, independent of its dispersion, dimensionality, or spectral density. Moreover, since the construction is formulated entirely at the single-particle level and does not depend on fermionic statistics, it extends equally to noninteracting bosonic baths. As a result, the Cayley-tree decomposition provides a unified, broadly applicable framework for representing noninteracting environments in quantum impurity models within a tree tensor network.

 \begin{figure}[t]
    \centering
    \includegraphics[]{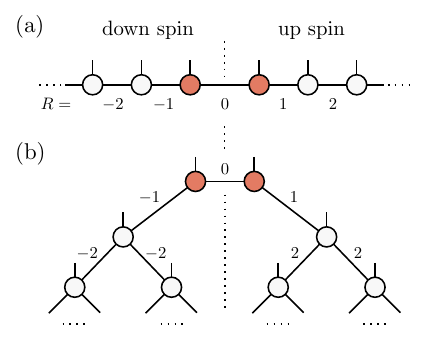}
    \caption{The SIAM in an MPS (a) and TTN (b) representation with explicit spin separation. The left and right branches correspond to the spin-up and spin-down channels, respectively. Orange nodes denote the impurity spins. The quantity \(|R|\) measures the minimal graph distance of a bath site from the impurity along the connecting path, with \(R=0\) assigned to the separation between the two impurity spins. Positive (negative) values of \(R\) label the coordinate (or layer) of spin-up (down) bath sites.
}
    \label{fig:tensor_reps}
\end{figure}

 The mapping is constructed iteratively. We first define the root fermion operator
\begin{equation}
  a_{1}=\sum_{k}\frac{V_k}{V}c_{k}, \quad V=\sqrt{\sum_{k}|V_{k}|^{2}} , \label{Eq:fermion1} 
\end{equation}
 which mediates the coupling between the impurity and the bath. In terms of \(a_1\), the hybridization Hamiltonian reduces to
\begin{equation}
  H_c = V \left( d^\dagger a_1 + a_1^\dagger d \right) .
\end{equation}

 The binary-tree representation of \(H_b\) is built starting from the root basis state in the first layer, which consists of a single site: $|t_{1}\rangle =a_1^\dagger|0\rangle$ with $|0\rangle$ the vacuum. Since \(|t_1\rangle\) is not orthogonal to the original bath basis
\(|k\rangle = c_k^\dagger |0\rangle\),
we generate an orthonormal basis set
\(\{ |t_i\rangle,\, i=1,\ldots,N \}\)
by applying the Gram-Schmidt reorthogonalization procedure, taking \(|t_1\rangle\) as the initial vector.

 As an explicit example, the second basis state \(|t_2\rangle\) is obtained by orthogonalizing an arbitrarily chosen state \(|k_1\rangle\)  in the bath,
\begin{eqnarray}
  |t_2\rangle
  &=& \alpha \left( |k_1\rangle - \langle t_1 | k_1 \rangle \, |t_1\rangle \right) \nonumber \\
  &=& \alpha \left( |k_1\rangle - v_{k_1} \, |t_1\rangle \right),
\end{eqnarray}
where the normalization constant \(\alpha\) is chosen such that \(\langle t_2 | t_2 \rangle = 1\).
The remaining \((N-2)\) basis states follow analogously.

The iterative construction proceeds as follows:

\vspace{2mm}

\noindent\textbf{(i)} In the basis $ ( |t_{1}\rangle ,|t_{2}\rangle ,\cdots ,|t_{N}\rangle ) $, the bath Hamiltonian $H_b$ can be represented as
\begin{eqnarray}
H_b = \left(
\begin{array}{cccc}
h_{11} & h_{12} & \cdots & h_{1N} \\
h_{21} & h_{22} & \cdots & h_{2N} \\
\vdots & \vdots & \ddots  & \vdots \\
h_{N1} & h_{N2} & \cdots & h_{NN} 
\end{array}
\right), 
\end{eqnarray}
with the matrix elements $h_{ij} = \langle t_i | H_b | t_j \rangle$. 

Separating the root state from the remaining subspace, we rewrite \(H_b\) as 
\begin{equation}
H_b =\left(
\begin{array}{cc}
h_{11} & S_1 \\
S_1^\dagger & H_1
\end{array}%
\right)
\end{equation}
 where $S_1=(h_{12} , h_{13} , \cdots  , h_{1N} ) $ is a $\left( N-1\right)$-dimensional vector and \(H_1\) is the \((N-1)\times(N-1)\) sub-Hamiltonian.

\vspace{2mm}

\noindent\textbf{(ii)} We diagonalize $H_1$ by a unitary matrix $U$
\begin{equation}
H_1 = U \lambda U^\dagger,
\end{equation}
with \(\lambda\) a diagonal matrix. The Hamiltonian then becomes
\begin{eqnarray}
H_b
&=&
\begin{pmatrix}
1 & 0 \\
0 & U
\end{pmatrix}
\begin{pmatrix}
h_{11} & S_1 U \\
U^\dagger S_1^\dagger & \lambda
\end{pmatrix}
\begin{pmatrix}
1 & 0 \\
0 & U^\dagger
\end{pmatrix}.
\end{eqnarray}

This defines a set of new basis states
\begin{equation}
(|\tilde t_2\rangle,\ldots,|\tilde t_N\rangle)
\equiv
(|t_2\rangle,\ldots,|t_N\rangle) U,
\end{equation}
in which the sub-Hamiltonian is diagonal, and the couplings between
\(|t_1\rangle\) and \(|\tilde t_m\rangle\) are encoded in
\begin{equation}
S_1 U = (\alpha_2,\alpha_3,\ldots,\alpha_N).
\end{equation}

Consequently,
\begin{eqnarray}
H_b |t_1\rangle
&=&
h_{11} |t_1\rangle
+ \sum_{m=2} \alpha_m |\tilde t_m\rangle, \label{Eq:Hbt1}
\\
\langle \tilde t_m | H_b | \tilde t_n \rangle
&=&
\lambda_m \delta_{mn},
\qquad (m,n\ge2 ).
\end{eqnarray}

\noindent\textbf{(iii)} 
We partition the states \(\{|\tilde t_m\rangle, m=2,\ldots,N\}\) into two subsets \(I_1\) and \(I_2\) of approximately equal size. For each subset, we define a branch root state,
\begin{eqnarray}
  |t_{11} \rangle  &=&\frac{1}{b_1}\sum_{m\in I_1}\alpha _{m}|\tilde{t}_{m}\rangle , \quad b_1 = \sqrt{\sum_{m\in I_1} \alpha_m^2} ,
\\
|t_{21} \rangle  &=&\frac{1}{b_2}\sum_{m\in I_2}\alpha _{m}|\tilde{t}_{m}\rangle ,  \quad b_2 = \sqrt{\sum_{m\in I_2} \alpha_m^2} . 
\end{eqnarray}
 The partition is chosen to minimize \(|b_1-b_2|\), ensuring a balanced branching. Eq.~\eqref{Eq:Hbt1} then reduces to
\begin{equation}
H_{b}|t_{1}\rangle =h_{11} |t_{1}\rangle +b_1|t_{11} \rangle
+b_2 |t_{21}\rangle .
\end{equation}

 The basis states within each branch are subsequently reorthogonalized using the Gram-Schmidt procedure, yielding orthonormal sets \( \{ |t_{1m}\rangle , (m=1, \cdots , N_1) \} \) and \( \{|t_{2m}\rangle , (m=1, \cdots , N_2)\} \), where $N_1$ and $N_2$ are the basis numbers in the corresponding subsets. The corresponding sub-Hamiltonians take the form
\begin{equation}
H_b^{\ell} =
\begin{pmatrix}
h^{\ell}_{11} & \cdots & h^{\ell}_{1N_\ell} \\
\vdots & \ddots & \vdots \\
h^{\ell}_{N_\ell 1} & \cdots & h^{\ell}_{N_\ell N_\ell}
\end{pmatrix},
\qquad \ell=1,2,
\end{equation}
with \(h^{\ell}_{ij}=\langle t_{\ell i}|H_b|t_{\ell j}\rangle\).

Collecting all terms, the bath Hamiltonian assumes the block form
\begin{equation}
H_b =
\begin{pmatrix}
h_{11} & V_1 & V_2 \\
V_1^\dagger & H_b^{1} & 0 \\
V_2^\dagger & 0 & H_b^{2}
\end{pmatrix},
\end{equation}
where
\(V_1=(b_1,0,\ldots,0)_{N_1}\)
and
\(V_2=(b_2,0,\ldots,0)_{N_2}\).

% \end{enumerate}

\vspace{2mm}

Steps \textbf{(i)} -\textbf{(iii)} complete one iteration of the binary branching. Repeating the procedure recursively on each sub-Hamiltonian generates a full Cayley-tree representation of the bath Hamiltonian. 

 In the present construction, a single root bath site, defined by Eq.~\eqref{Eq:fermion1}, is coupled to the impurity, ensuring local equivalence to the conventional chain decomposition. More generally, the bath basis states ${|k\rangle, k=1,\cdots,N}$ can be divided into $n$ subsets, each of which is mapped onto a Cayley tree following the same procedure. In this way, the impurity is coupled directly to $2n$ Cayley trees, since each subset generates two such trees. This extension provides a more flexible representation of the bath and can potentially lead to a more efficient encoding of entanglement. A notable special case is $n=2$, where one subset contains all occupied states $(\epsilon_k<0)$ and the other all unoccupied states $(\epsilon_k>0)$. Such a bipartite partition of the bath was previously adopted in the MPS study of the SIAM in Ref.~\cite{Kohn2021}.

 In the SIAM, there is no hybridization between different spin channels, since the Hubbard interaction is purely local. As a result, the up-spin and down-spin channels can be treated independently and are coupled only at the impurity site, as illustrated in Fig.~\ref{fig:tensor_reps}. This decoupling reduces the local physical dimension at each bath site from 4 to 2, yielding a substantial computational advantage and enabling a more accurate representation of the quantum state at reduced cost.

 Our approach is distinct from a previously proposed tree-lattice-based impurity solver introduced in Ref.~\cite{PhysRevB.104.115119}. In that work, a tree structure is used to interconnect multiple impurity orbitals, while the bath degrees of freedom are encoded in a one-dimensional chain geometry. By contrast, in our method, the bath is mapped onto a Cayley tree, providing a more natural and efficient framework for capturing the multiscale entanglement between the impurity and the bath. Moreover, our architecture is inherently extensible: in multi-impurity problems, it can be directly combined with auxiliary tree structures connecting the impurities, providing a unified and powerful representation for complex quantum impurity models.

\subsection{Calculation of the Green's function}
\label{Sec:TTN}

Since both MPS and TTN~\cite{Shi2006} states can always be brought into canonical forms, we compute the ground state \( |\Psi_0\rangle \) and its energy \( E_0 \) using the simple-update method~\cite{Jiang2008,Li2012,Chen2025}. This approach enables efficient and numerically stable optimization of the ground-state wavefunction, although the local updates are not strictly optimal at each iteration.

 To compute dynamical properties, we first evaluate the real-time evolution of the impurity Green’s function using the time-evolving block decimation (TEBD) method~\cite{Vidal2004}, implemented with a first-order Trotter-Suzuki decomposition. Within the tree tensor network framework, the real-time Green’s function can be computed directly, providing direct access to dynamical information on the real-frequency axis. To enhance the frequency resolution, the resulting finite-time data are extrapolated using the linear prediction (LP) technique~\cite{white2004, Barthel2009,Wolf2014_2,Ganahl2014} prior to Fourier transformation, which yields the retarded Green’s function and the corresponding spectral function. This approach avoids the ill-conditioned analytic continuation procedures required in imaginary-time or Matsubara-frequency methods, enabling a controlled and accurate determination of real-frequency spectral properties.

One may also use the time-dependent variational principle (TDVP) to compute the real-time evolution. Unlike TEBD, TDVP is free of Trotter errors. However, it requires sequential forward and backward sweeps to construct and update the environment tensors, making the algorithm computationally demanding. By contrast, TEBD with simple updates involves only local operations and avoids these costly environmental contractions, rendering it highly efficient for parallel implementation. For the binary tree lattice considered here, for example, the bonds can be partitioned into three non-overlapping sets, such that within each set all time-evolution gates commute and can therefore be applied simultaneously to the tree tensor network. As the Trotter error decreases with decreasing time step, TEBD thus provides a favorable balance between accuracy and computational efficiency.

For the SIAM, we calculate the retarded impurity Green’s function defined as
\begin{eqnarray}
    G(t) &=& -i\theta(t)\langle \Psi_0|\{d(t),d^\dagger(0)\}|\Psi_0 \rangle  \nonumber \\
    &=& -i\theta(t) \left[ G_+ (t)+G_- (t) \right] \label{eq:Gt}
\end{eqnarray}
where $\theta(t)$ is the step function and $|\Psi_0\rangle$ denotes the ground state. The greater and lesser Green’s functions are defined by
\begin{eqnarray}
     G_+(t) &=& \langle \Psi_0|de^{-iHt}d^\dagger|\Psi_0 \rangle e^{iE_0t},\\
    G_-(t) &=& \langle \Psi_0|d^\dagger e^{iHt}d|\Psi_0 \rangle e^{-iE_0t} . 
\end{eqnarray}
 For a particle–hole symmetric ground state, $G_+(t)=G^*_-(t)$, Eq.~\eqref{eq:Gt} reduces to $G(t) = -2i\theta(t)\text{Re}G_+(t)$. 

 The retarded Green’s function in the frequency domain, \( G(\omega) \), is obtained via Fourier transformation of \( G(t) \). The spectral function is then determined from the imaginary part of $G(\omega)$
\begin{equation}
    A(\omega) = -\frac{1}{\pi}\text{Im} G(\omega) .
\end{equation}

\begin{figure}[t]
    \centering
    \includegraphics[width=0.9\linewidth]{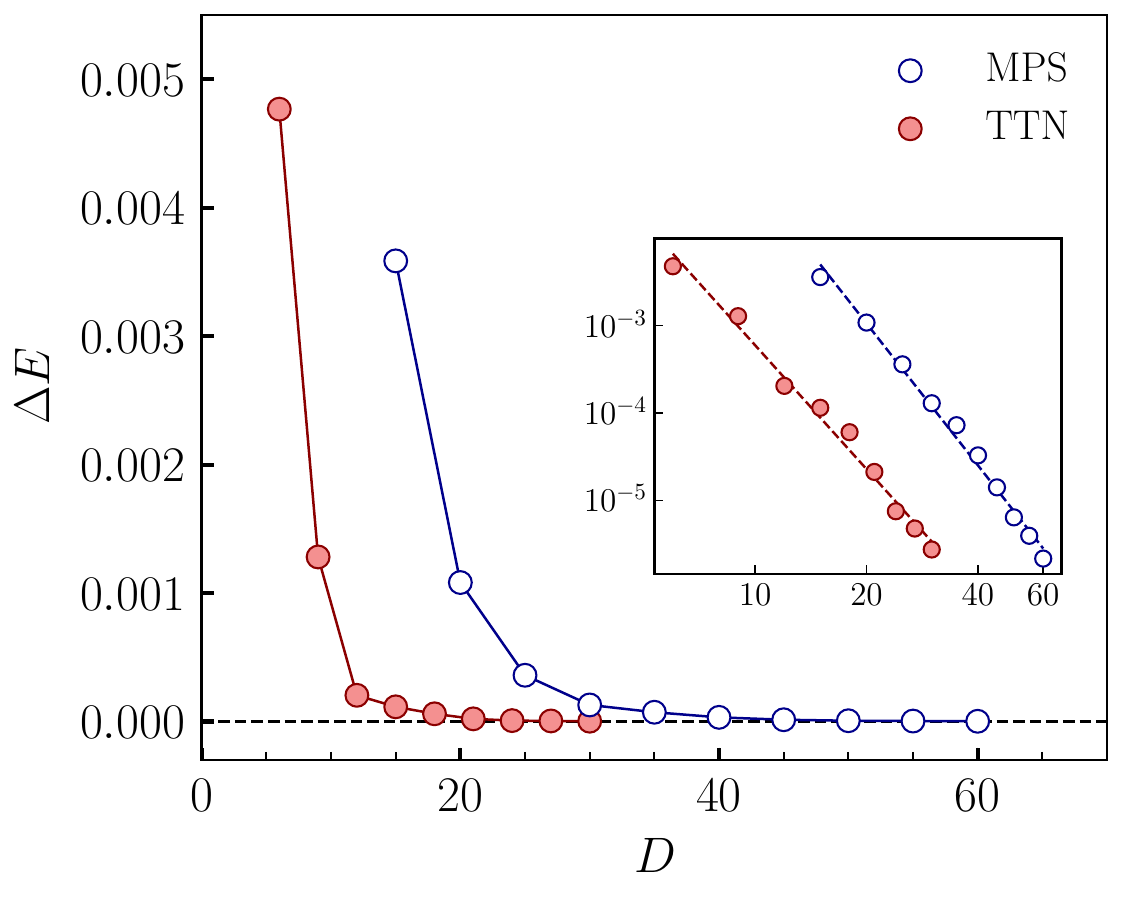}
    \caption{Ground-state energy convergence. Ground state energy difference, $\Delta E=E(D)-E_0$, as a function of the bond dimension $D$. The reference energy $E_0$ is obtained by extrapolating the DMRG results to the $D\to\infty$ limit. The inset shows the same data on a log–log scale. Dashed lines denote power-law fits of the form $\Delta E = \alpha D^{-\beta}$, where $(\alpha, \beta) = (11048, 5.39)$ for the MPS results and $(\alpha, \beta) = (31, 4.71)$ for the TTN results.
    }
    \label{fig:ground_energy}
\end{figure}

\subsection{Model parameters}

 In impurity models, and more generally within the framework of DMFT, the fermionic bath and its coupling to the impurity are fully characterized by the hybridization function $\Delta(\omega)$. This function encodes all bath degrees of freedom relevant to the impurity and is related to the bath parameters $\{\varepsilon_k, V_k\}$ through
\begin{equation}
    \Delta(\omega) = \pi \sum_k |V_k|^2\,\delta(\omega-\varepsilon_k).
\end{equation}

 Conversely, for a given $\Delta(\omega)$, the bath parameters can be obtained using a deterministic discretization scheme~\cite{Vega2015,Bauernfeind2017},
\begin{eqnarray}
    V_k^2 &=& \frac{1}{\pi}\int_{I_k} d\omega\, \Delta(\omega), \\
    \varepsilon_k &=& \frac{1}{\pi V_k^2}\int_{I_k} d\omega\, \omega\,\Delta(\omega),
    \label{Eq:bath_disc}
\end{eqnarray}
 where the interval $I_k$ is associated with a single bath orbital. This construction replaces the continuum bath by discrete levels with weights $V_k^2$ centered at $\varepsilon_k$, while preserving essential spectral sum rules~\cite{PhysRevB.78.115102}. 

 A number of discretization schemes exist for representing a continuous hybridization function with a finite bath of size $N$. Common choices include linear discretization, frequently employed in DMRG calculations~\cite{Ganahl2015}, and logarithmic discretization, characteristic of NRG methods~\cite{Bulla2008}. For tensor network simulations, numerical stability is enhanced by adopting a scheme that yields approximately constant coupling strengths $V_k$ across intervals, rather than constant $I_k$~\cite{Bauernfeind2017}.
 
 As a benchmark, we consider the particle-hole symmetric case with a constant hybridization function \( \Delta(\omega)=\Delta_0=0.1 \), local interaction \(U=0.8\), and impurity level \(\varepsilon_d=-U/2=-0.4\). Unless otherwise specified, the bath is discretized into $N=63$ sites within the bandwidth [-1,1]. We solve the SIAM using the TTN solver and assess its performance by comparing it with MPS calculations. Unless stated otherwise, the MPS results presented in this section are obtained using the conventional chain mapping.

\begin{figure}[t]
    \centering
    \includegraphics[width=0.85\linewidth]{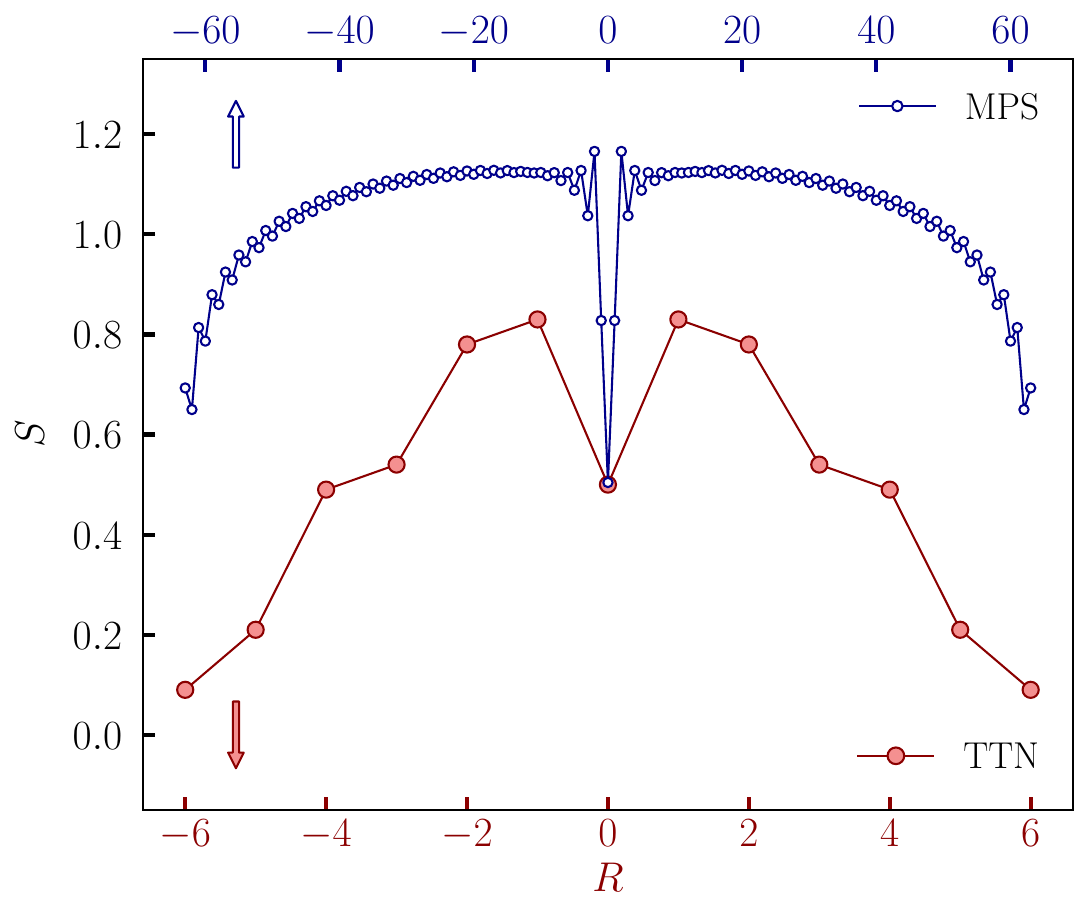}
    \caption{Entanglement entropy $S$ as a function of the distance $R$ from the impurity site for MPS with $D=60$ (blue circles) and TTN with $D=30$ (red circles). For the TTN, the data points represent averages over all bonds at the same distance $R$.
    }
    \label{fig:mps_tree_entanglement}
\end{figure}

\section{Results}
\label{Sec:results}

\begin{figure}[t]
    \centering
    \includegraphics[width=0.85\linewidth]{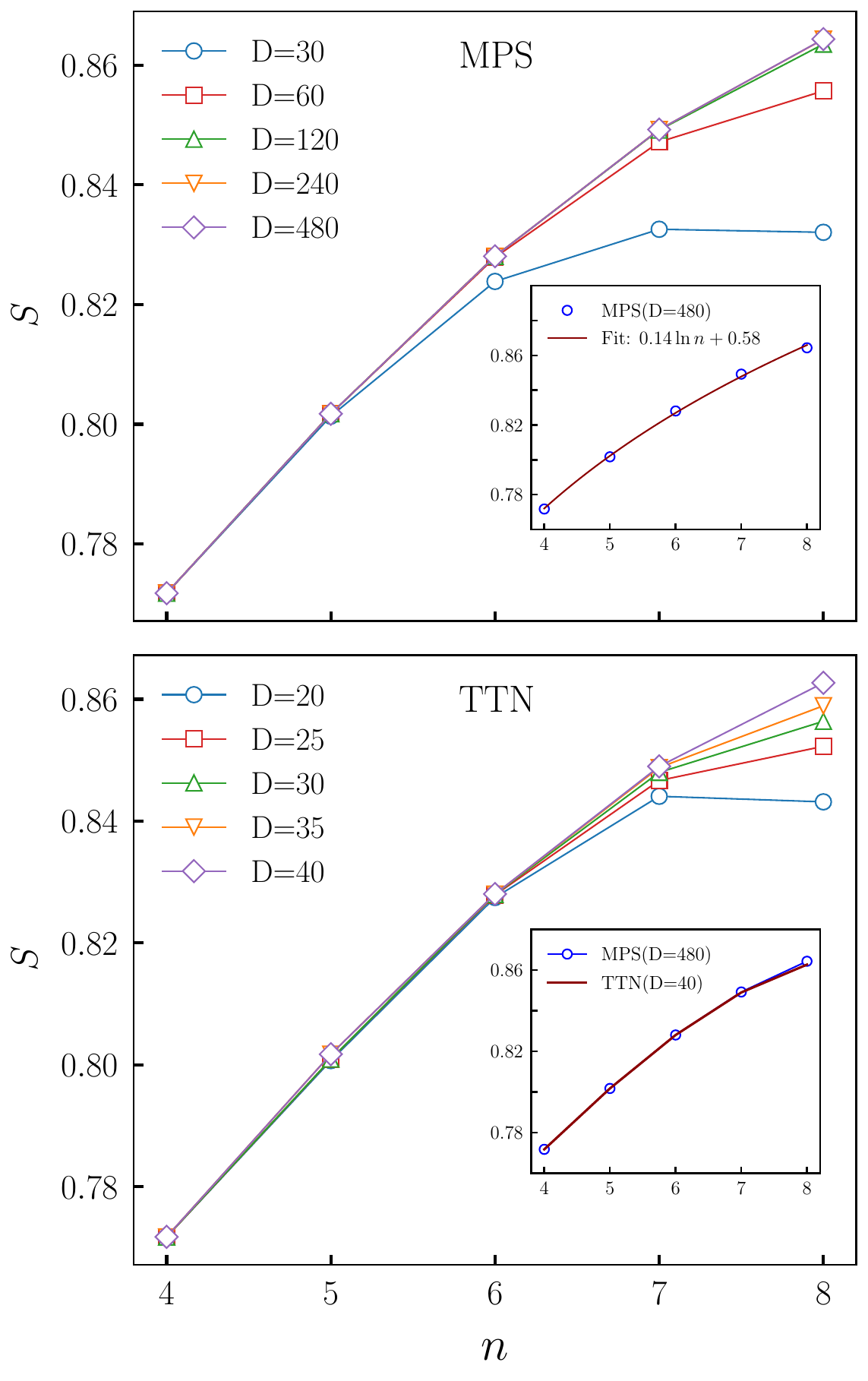}
    \caption{Entanglement entropy $S$ at the bond adjacent to the impurity ($R=1$) as a function of the logarithmic system size $n=\log_2 N$. The top and bottom panels show the convergence of $S$ with bond dimension $D$ for the MPS and TTN calculations, respectively. The inset of the upper panel shows a logarithmic fit to the converged MPS data ($D=480$), yielding $S \approx 0.14 \ln n + 0.58$. The inset of the lower panel compares MPS ($D=480$) and TTN ($D=40$) results and demonstrates quantitative agreement.
    }
    \label{fig:S_vs_n}
\end{figure}

\begin{figure}[t]
    \centering
    \includegraphics[width=0.85\linewidth]{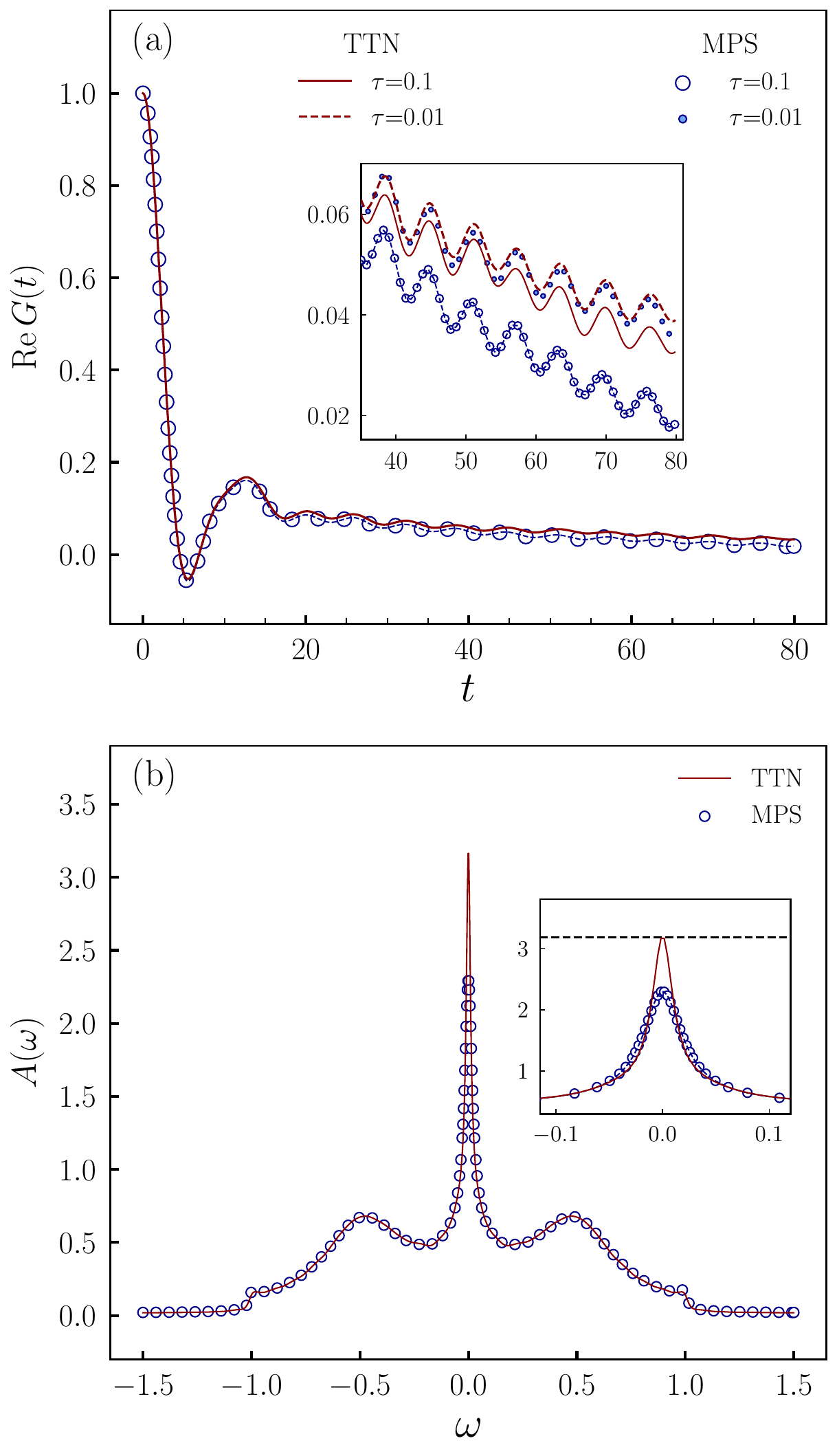}
    \caption{(a) Time evolution of $\mathrm{Re}\,G(t)$ obtained using TTN ($D=30$) and MPS ($D=90$) with time steps \(\tau=0.1\) and \(\tau=0.01\). The two methods agree at small $t$, while deviations appear at large $t$ (inset). (b) Spectral function \(A(\omega)\) obtained from the Fourier transform of \(G(t)\). The inset highlights the discrepancy near \(\omega=0\). The exact value at \(\omega=0\), \(A(0)=1/(\pi\Delta_0)\approx3.183\), is shown by the black dashed line. 
    }
    \label{fig:Re_Gt}
\end{figure}

\subsection{Ground state properties}
\label{Sec:ground_state}

 Figure~\ref{fig:ground_energy} illustrates the convergence of the ground-state energy obtained using MPS and TTN as a function of the bond dimension \(D\). The reference energy \(E_0\) is taken from DMRG calculations extrapolated to the $D\to \infty$ limit. In comparison, the TTN results exhibit a faster convergence of the ground-state energy with increasing \(D\) than those obtained using MPS. Both methods exhibit power-law convergence of the form $E(D) \approx E_0 + \alpha D^{-\beta}$. The fitted exponent is $\beta = 5.39$ for MPS and $\beta = 4.71$ for TTN. The larger exponent for MPS indicates a faster asymptotic convergence at large $D$. However, the prefactor $\alpha$ is significantly smaller for TTN, implying that TTN achieves higher accuracy in the low- to intermediate-bond-dimension regime. The two error curves intersect at a crossover bond dimension of approximately $D_c \sim 6000$, which is far beyond the range relevant for practical calculations. This behavior demonstrates that TTN is a more efficient impurity solver than MPS in practice.

 To further elucidate the origin of TTN's efficiency, we analyze the distribution of entanglement entropy in the ground state. For a fair comparison, we choose bond dimensions that yield comparable ground-state energy accuracy: $D = 60$ for MPS and $D = 30$ for TTN. The corresponding entanglement profiles are shown in Fig.~\ref{fig:mps_tree_entanglement}. The horizontal axis represents the graph distance $R$ from the impurity site (see Fig.~\ref{fig:tensor_reps}). Owing to the branching structure of the TTN, multiple bonds correspond to the same distance $R$. We plot the entanglement entropy averaged over all TTN bonds at a given $R$.

  Both profiles exhibit a characteristic ``M-shaped'' structure arising from the spin-decoupled representation. The local minimum at \(R=0\) reflects the relatively weak spin correlations on the impurity site. For bonds at a layer distance \(R=\pm 1\), the entanglement entropy obtained with the TTN equals that of the MPS. This local equivalence arises because the tensor-network connectivity for these bonds is the same in both geometries, as illustrated in Fig.~\ref{fig:tensor_reps}. Clear differences emerge in layers farther away from the impurity. In the MPS, the entanglement entropy remains close to its maximum value allowed by the bond dimension across nearly all bonds. In contrast, the TTN exhibits an approximately linear decay of entanglement entropy with increasing layer separation, reflecting its intrinsic hierarchical structure. As a result, the maximum entanglement entropy in the TTN is substantially reduced, highlighting its efficiency in capturing the intrinsic entanglement structure of the ground state.

 Figure~\ref{fig:S_vs_n} shows the entanglement entropy $S$ at $R=1$ as a function of the logarithmic bath size $n=\log_2 N$. In contrast to the linear scaling $S\propto n$ characteristic of bulk critical systems predicted by conformal field theory (CFT)~\cite{Pasquale2004, Hastings_2007, Eisert2010, Laflorencie_2016}, we observe a slower growth, $S\sim \ln n \sim \ln ( \ln N )$. This logarithmic dependence indicates that the system remains critical, with gapless excitations at the Fermi level~\cite{Affleck1991, Sørensen_2007, Affleck_2009, Eriksson2011}. Although boundary CFT predicts that the impurity contribution to the entanglement entropy ultimately saturates in the thermodynamic limit, this $\ln n$ growth suggests that the system size $N$ is comparable to or smaller than the spatial extent of the Kondo screening cloud, $\xi_K\sim v_F/T_K$~\cite{Affleck_2009,Eriksson2011}. The resulting curvature of $S$ thus encodes the gradual refinement of the bath representation and the progressive buildup of long-range correlations as the system slowly moves toward the Kondo fixed point~\cite{wilson1975,Affleck_2009}.

As shown in the inset of the lower panel, the TTN faithfully reproduces the logarithmic dependence of the entanglement entropy $S$ on $n$ and attains quantitative agreement with MPS results obtained at a bond dimension of $D\simeq 480$, while requiring only $D\simeq 40$. This order-of-magnitude reduction in bond dimension directly demonstrates that the TTN geometry allocates entanglement resources in a manner optimally matched to the multiscale structure of the impurity–bath correlations, yielding a substantially more efficient representation of the quantum impurity problem.

 At a given layer, the computational cost per local tensor update scales as \(O(D^4)\) for the TTN and \(O(D^3)\) for the MPS. In the MPS geometry, the entanglement entropy is nearly uniform across the lattice, apart from a few sites near the boundaries, so a large bond dimension is required on almost all bonds. Consequently, the total computational cost scales approximately as \(O(ND^3)\).

 By contrast, in the TTN, the entanglement entropy decreases rapidly with increasing layer distance \(|R|\). As a result, only a small subset of bonds in the layers closest to the root (\(R \leq 2\)) requires a large bond dimension. The outermost layer, which contains nearly half of all sites, has bond dimensions of at most \(D=2\), while the second outermost layer, accounting for roughly one quarter of all sites, also exhibits bounded entanglement with bond dimensions limited to \(D \leq 8\). Owing to this intrinsic multiscale structure, the total computational cost scales as \(O(N_0 D^4)\), where \(N_0 \ll N\) denotes the number of bonds in the first few layers that carry the maximal entanglement entropy.

 Furthermore, the tree geometry combined with the simple-update scheme~\cite{Jiang2008,Li2012,Chen2025} is naturally amenable to parallelization, in sharp contrast to the largely sequential nature of DMRG-based algorithms. Taking into account that the TTN attains comparable accuracy with a significantly smaller bond dimension than the MPS, these considerations demonstrate that the TTN can capture impurity physics at a substantially reduced computational cost.

The efficiency of the TTN impurity solver becomes clear when compared with other established approaches. Unlike CTQMC~\cite{Rubtsov2005,Werner2006,Gull2011}, which accesses ground-state properties only indirectly via low-temperature extrapolation, the TTN computes the ground state directly. Its power-law convergence of the ground-state energy with bond dimension parallels that of ED–based solvers with increasing bath size, but without the exponential growth of the Hilbert space or the severe finite-bath discretization errors inherent to ED~\cite{Caffarel1994,Capone2007,Granath2012,Lu2014}. While the NRG~\cite{Bulla1999,Bulla2001,Bulla2008} excels at low-energy physics but compromises high-energy resolution due to logarithmic discretization, the TTN employs linear or optimized discretization schemes, providing a uniform description across the full energy spectrum. Compared to the NORG~\cite{He2014}, whose performance can degrade in challenging regimes such as half filling due to basis optimization issues, the TTN does not rely on a specific single-particle basis, offering a robust framework for accurate ground-state calculations.

\subsection{Spectral function}

 To obtain the spectral function, we compute the real-time evolution of the retarded Green’s function $G(t)$ using both TTN and MPS with different Trotter time steps $\tau$. Bond dimensions $D=30$ for TTN and $D=90$ for MPS are employed, chosen to ensure comparable levels of convergence, as established by the bond-dimension analysis in Appendix~\ref{sec:real_time_bond_dimes}. As shown in Fig.~\ref{fig:Re_Gt}(a), the two methods produce nearly identical results at short times, where the entanglement growth remains limited. At longer times, however, a systematic deviation develops, most clearly visible in the inset, reflecting the increasing importance of bath-induced correlations. Despite using a substantially smaller bond dimension, the TTN maintains an accuracy comparable to, and in the long-time regime superior to, that of the MPS.

 These results reaffirm the previous observation that the structure of the TTN facilitates a more efficient representation of bath entanglement at finite bond dimensions, leading to a more accurate description of long-range correlations in the time evolution. In addition, the TTN exhibits a smaller prefactor in the Trotter error. Although the leading Trotter error scales as $O(\tau^2)$ for both geometries, the MPS results show a pronounced dependence on the time step, with visible improvement when $\tau$ is reduced from $0.1$ to $0.01$, whereas the TTN results remain largely unchanged. The Trotter error is controlled by the product of $\tau$ and the hopping constant at each bond. Because the hopping constants decay much faster in the Cayley-tree mapping than in the 1D chain mapping, the prefactor of the Trotter error is correspondingly reduced. This geometric suppression of the Trotter error is explicitly verified using an exact free-fermion model in Appendix~\ref{sec:appendix_trotter}.
This reduced sensitivity to the Trotter step allows the TTN to achieve high accuracy at relatively large $\tau$, providing a significant computational advantage for applications in DMFT calculations, where repeated real-time evolutions are required within self-consistency loops.

 To obtain high-resolution spectral functions, the real-time data are extended using linear prediction (LP) with parameters $(N_p,\delta)=(400,10^{-7})$, adding 8000 additional time points prior to Fourier transformation (see Appendix~\ref{Sec:LP} for details). The resulting spectral functions $A(\omega)$ obtained with $\tau=0.1$ are shown in Fig.~\ref{fig:Re_Gt}(b).

 The overall spectral line shapes obtained from TTN and MPS are in close agreement over a broad frequency range. However, a clear discrepancy emerges in the low-frequency regime near $\omega=0$, as highlighted in the inset. For a constant hybridization function $\Delta_0$, the exact zero-frequency spectral weight at particle–hole symmetry and zero temperature is given by $A(0)=1/(\pi\Delta_0)\approx3.183$ for the parameters used here~\cite{Langreth1966,Bulla2008}. The TTN reproduces this value with a relative error of approximately $6\times10^{-3}$, which is significantly smaller than that obtained from the MPS calculation.

\subsection{Entanglement evolution}

\begin{figure}[t]
    \centering
    \includegraphics[width=0.85\linewidth]{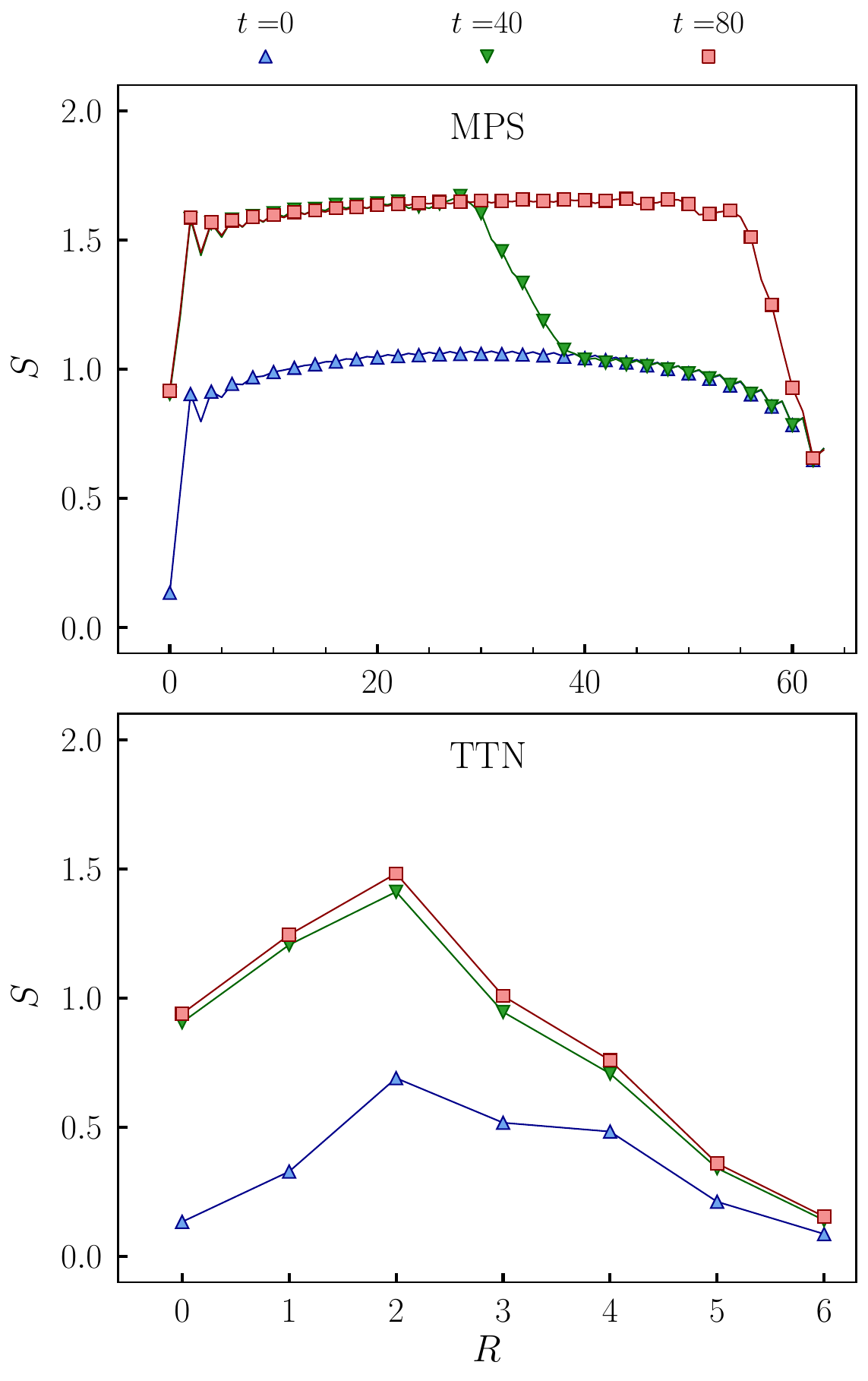}
    \caption{ 
    Time evolution of the entanglement entropy profile at three representative times $t=(0, 40, 80)$ for MPS with $D=90$ and TTN with $D=30$.}
    \label{fig:S_evolution_sparse}
\end{figure}

 As demonstrated in Fig.~\ref{fig:mps_tree_entanglement}, the TTN accurately captures the ground-state entanglement structure. We now show that this advantage extends to real-time dynamics.

 % Figure~\ref{fig:S_evolution_sparse} compares the entanglement entropy profiles at the initial time ($t=0$)and at $t=80$ 
 Figure~\ref{fig:S_evolution_sparse} compares the entanglement entropy profiles at three representative times $t=(0,40,80)$ (see Appendix~\ref{sec:real_time_S_dense} for the full time evolution). In the MPS evolution [Fig.~\ref{fig:S_evolution_sparse}(a)], entanglement rapidly saturates on nearly all bonds, quickly reaching the limit set by the finite bond dimension. 
 
 In contrast, the TTN exhibits structured entanglement evolution [Fig.~\ref{fig:S_evolution_sparse}(b)] that closely mirrors its ground-state behavior. Entanglement concentrates in the intermediate layers around $R\sim\pm 2$, while bonds in the innermost and outermost layers remain weakly entangled. This hierarchical distribution enables targeted increases in bond dimension only where needed, in sharp contrast to MPS, which requires nearly uniform increases across the entire system. Consequently, the TTN preserves higher fidelity during long-time dynamical evolution.

\begin{figure}[t]
    \centering
    \includegraphics[width=0.85\linewidth]{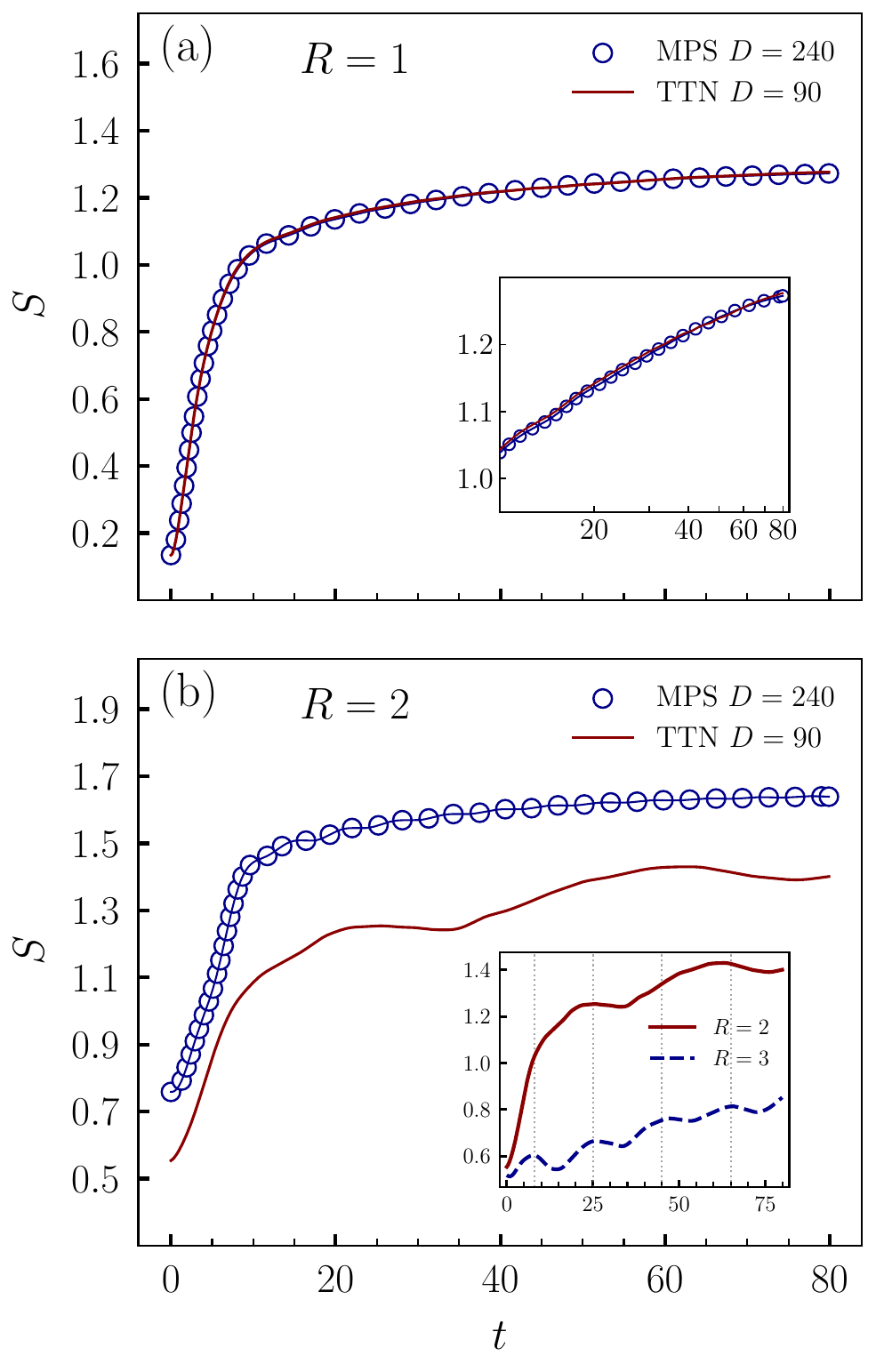}
    \caption{ 
    Time evolution of the entanglement entropy $S$ at (a) $R=1$ and (b) $R=2$. The inset in (a) shows $S$ plotted on a logarithmic time scale for intermediate times $t>10$. The inset in (b) compares the TTN results for $S$ at $R=2$ and $R=3$ with $D=90$. Vertical dotted lines indicate characteristic timescales associated with the oscillatory behavior.
    }
    \label{fig:S_t_differentR}
\end{figure}

 Figure~\ref{fig:S_t_differentR} shows the time evolution of the entanglement entropy at $R=1$ and $R=2$. The MPS and TTN calculations use bond dimensions $D=240$ and $D=90$, respectively (see Appendix~\ref{sec:real_time_S_dense} for details on the convergence of $S$ with bond dimension). At $R=1$, the MPS and TTN have the same local geometry, so their entanglement dynamics are expected to be the same. This is confirmed by the results in Fig.~\ref{fig:S_t_differentR}. At short times, the entanglement entropy increases linearly, $S\sim t$, reflecting the ballistic spreading of correlations within the Lieb–Robinson light cone~\cite{Lieb1972, Calabrese2005}. At the intermediate time scale (the inset of Fig.~\ref{fig:S_t_differentR}(a)), $t\gtrsim 10$, the growth crosses over to a slower logarithmic behavior, $S\sim \ln t$. Such logarithmic scaling behavior is characteristic of gapless critical systems with a diverging correlation length, consistent with conformal field theory predictions~\cite{Calabrese2005, Sørensen_2007}.

 At $R=2$, the MPS entanglement entropy remains qualitatively similar to the behavior observed at $R=1$. In contrast, the TTN result shows a reduced overall magnitude and develops clear non-monotonic oscillations. These oscillations originate from the branching structure of the TTN, which divides the bath into finite sub-branches. Excitations propagating along these branches undergo partial reflections at their boundaries, leading to oscillatory entanglement dynamics. This effect becomes more apparent when comparing the TTN results at $R=2$ and $R=3$, as shown in the inset of Fig.~\ref{fig:S_t_differentR}(b). At $R=3$, the oscillations are especially pronounced and display an approximate period of $T\simeq 20$, while at $R=2$ the period is roughly twice as long. This scaling of the oscillation period reflects the finite extent of the sub-bath geometry inherent to the TTN~\cite{Vega2015}.

\subsection{Energy-split multi-branch TTN}
\label{sec:multi-branch}

\begin{figure}[t]
    \centering
    \includegraphics[width=0.85\linewidth]{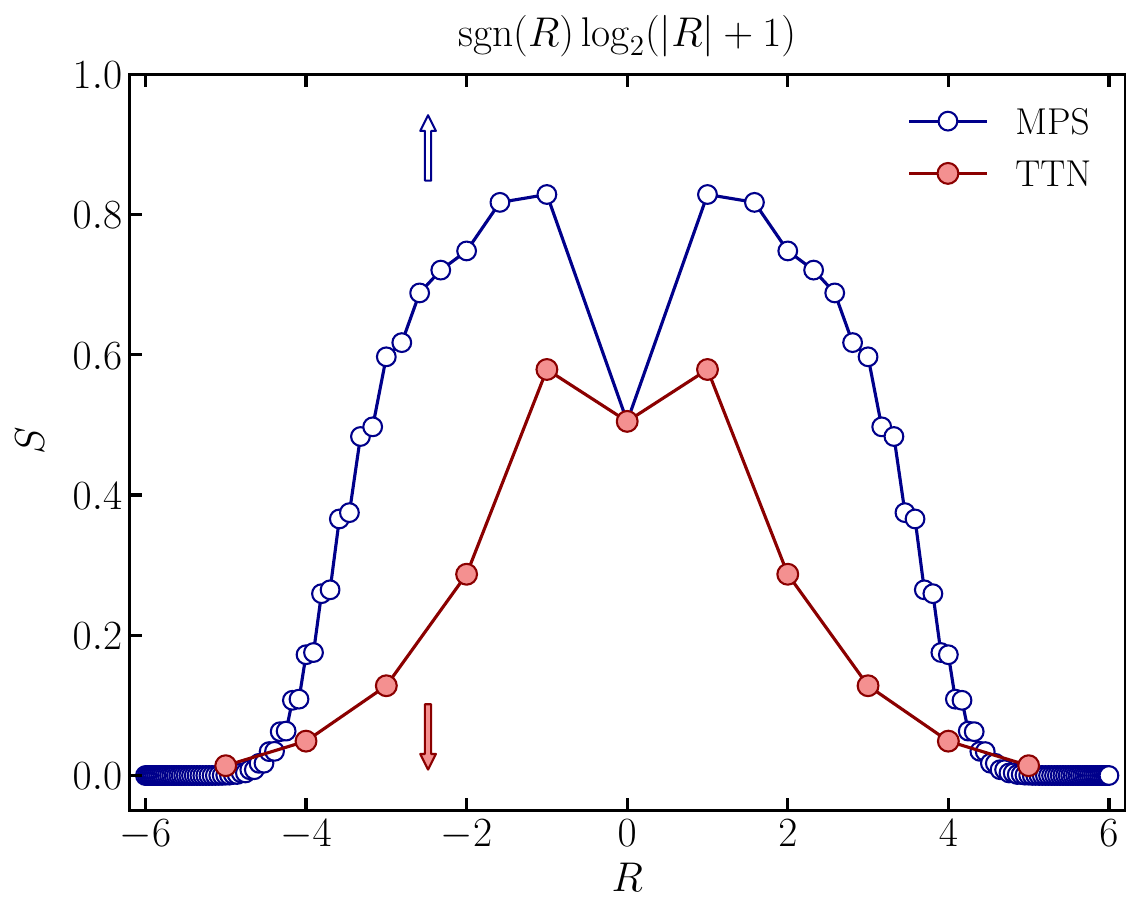}
    \caption{Comparison of the ground-state entanglement entropy profiles obtained with the energy-split TTN (red circles, $D=30$) and the interleaving-energy-split MPS (blue circles, $D=60$). To reflect the hierarchical structure of the TTN, the horizontal axis for the MPS data is rescaled as $\mathrm{sgn} (R)\log_2(|R|+1)$.}
    \label{fig:mps_ttn_compare_E_sep}
\end{figure}

As mentioned in Sec.~\ref{Sec:tree_transform}, the bath can be separated into two or more subsets, each mapped to an independent binary Cayley tree that couples to the impurity via its root site. A particularly natural partition is to divide the bath into occupied and unoccupied bath orbitals prior to coupling to the impurity. This branch construction was first introduced by Kohn and Santoro in their interleaving-energy-split MPS treatment of SIAM~\cite{Kohn2021}.

In our framework, this idea corresponds to an energy-split, multi-branch TTN, in which the impurity couples to two distinct root-bath sites in each spin channel. At half filling, the $N=63$ bath is divided into 32 initially occupied orbitals and 31 initially unoccupied orbitals. The full system, including the impurity, therefore contains 32 spin-up and 32 spin-down electrons.

 Figure~\ref{fig:mps_ttn_compare_E_sep} compares the ground-state entanglement entropy profiles obtained using the energy-split multi-branch TTN and the interleaving-energy-split MPS approach. To account for the hierarchical structure of the bath vectors, the MPS results are plotted against $\log_2(|R|+1)$. For both ansatze, separating bath states by occupation significantly reduces the entanglement entropy relative to calculations without such separation. Moreover, the entanglement entropy decreases rapidly with increasing bond distance $R$ from the impurity, becoming nearly zero at large-$R$. 

  Although the interleaving-energy-split MPS already provides a substantial improvement over the standard Wilson chain mapping, the energy-split TTN still performs better. The maximal entanglement entropy at the bonds close to the impurity is consistently lower in the TTN than in the optimized MPS. Furthermore, the entanglement in the TTN decays more rapidly with the network depth. This further demonstrates that the tree geometry provides a more efficient representation of the hierarchical bath entanglement structure.

 Because the energy-split multi-branch TTN significantly reduces the entanglement entropy, spectral functions and other observables can be obtained with a smaller bond dimension $D$ while retaining the same accuracy as in Fig.~\ref{fig:Re_Gt}. This highlights the practical advantage of the multi-branch mapping for solving the impurity problem. At half filling, partitioning the bath according to occupation is a natural choice. More generally, however, for fillings away from half filling, such as one-third filling, how to partition the bath to minimize the entanglement structure remains an open question and deserves further investigation.

\section{Summary}
\label{Sec:summary}

In this work, we have introduced a TTN impurity solver based on a hierarchical Cayley-tree decomposition of the bath Hamiltonian. By explicitly mapping the noninteracting bath onto a tree geometry, the method provides a natural tensor-network formulation that directly reflects the multiscale entanglement structure inherent to quantum impurity problems. This representation fundamentally differs from conventional chain-based mappings and enables a more efficient distribution of entanglement across the network.
The Cayley-tree architecture naturally supports multi-branch constructions, for example, by explicitly separating the bath into occupied and unoccupied energy sectors. This energy-split strategy can further dilute the underlying multiscale correlations and yields a more efficient representation of the impurity problem.

Benchmark calculations for the SIAM show that the TTN solver is numerically more efficient than MPS-based methods. The Kondo ground state of the SIAM is critical~\cite{Affleck1991,Sørensen_2007,Affleck_2009,Eriksson2011}, characterized by a gapless resonance and a diverging correlation length. This criticality arises from a slow finite-size crossover associated with the gradual formation of the Kondo screening cloud. The resulting entanglement is long-ranged and highly inhomogeneous, making it costly to represent efficiently with MPS. 

The TTN gains its advantage from the multiscale nature of the Kondo problem. It captures strong entanglement near the impurity in the inner layers of the tree while describing weakly entangled bath degrees of freedom at large distances with minimal resources. This separation of scales is also evident in the real-time dynamics, where the entanglement entropy grows linearly at short times, $S\sim t$, within the Lieb–Robinson light cone~\cite{Lieb1972, Calabrese2005}, crosses over to a logarithmic increase, $S\sim \ln t$, at intermediate times~\cite{Calabrese2005, Sørensen_2007}, and eventually saturates at long times. 

In real-time evolution, the TTN preserves accurate long-time dynamics at smaller bond dimensions and shows reduced sensitivity to Trotter errors, owing to its hierarchical entanglement structure. As a result, computational resources can be used more efficiently by increasing the bond dimension only in the intermediate layers where entanglement is most pronounced.

Compared with other impurity solvers, the TTN approach avoids the analytic continuation required in CTQMC, circumvents the finite-bath discretization errors intrinsic to ED, provides a more uniform treatment of energy scales than NRG, and overcomes the active-orbital restrictions inherent to NORG. These features render the TTN particularly well suited for resolving low-energy dynamics in strongly correlated systems.

The TTN accurately reproduces the Kondo resonance height while simultaneously capturing the smooth spectral background characteristic of the true continuum limit. Its accuracy is governed by the bond dimension rather than by an explicit separation of energy scales. This enables a balanced description across the entire frequency range, from the sharp zero-frequency Kondo resonance to high-energy excitations. 

The framework developed in this work admits natural extensions in two directions. First, the TTN solver can be straightforwardly generalized to multi-impurity problems, where the entanglement structure becomes increasingly complex. Second, although the present study focuses on real-time evolution, the same formalism can be directly adapted to imaginary-time calculations. By combining balanced accuracy, computational efficiency, and extensibility, the TTN solver thus provides a versatile and powerful platform for the study of strongly correlated systems.

Several promising directions remain for future development. Beyond the binary tree used here, the tree geometry itself may be optimized by allowing variable branching across different layers, which could further reduce entanglement per bond and improve efficiency~\cite{Hikihara2023}. Identifying problem-adapted tree architectures remains an open question. In addition, the loop-free, branching structure of TTN is well suited for parallel implementations on modern GPU architectures~\cite{Milsted2019}. Together with possible integration into sampling-based variational Monte Carlo approaches~\cite{Ferris_2012,Cheng2019}, these developments may substantially accelerate impurity solvers and DMFT self-consistency cycles, enabling high-throughput studies of complex multi-orbital and cluster impurity models.

\section*{DATA AVAILABILITY}
The numerical simulation data that support the findings of this study are openly available in Zenodo at Ref.~\cite{zenodo_dataset}.

\section*{Acknowledgments}
This work is supported by the National Natural Science Foundation of China (12488201) and the National
Key Research and Development Project of China (2021ZD0301800).

\appendix

\section{Abbreviations}
\begin{enumerate}
    \item CTQMC: continuous-time quantum Monte Carlo
    \item CFT: conformal field theory
    \item DMFT: dynamical mean-field theory
    \item DMRG: density-matrix renormalization group
    \item ED: exact diagonalization
    \item LP: linear prediction
    \item MPS: matrix product state
    \item NRG: numerical renormalization group
    \item NORG: natural orbital renormalization group
    \item SIAM: single-impurity Anderson model
    \item TEBD: time-evolving block decimation
    \item TTN: tree tensor network
\end{enumerate}

\section{Convergence of the Green's function with bond dimension}
\label{sec:real_time_bond_dimes}

\begin{figure}[t]
    \centering
    \includegraphics[width=0.9\linewidth]{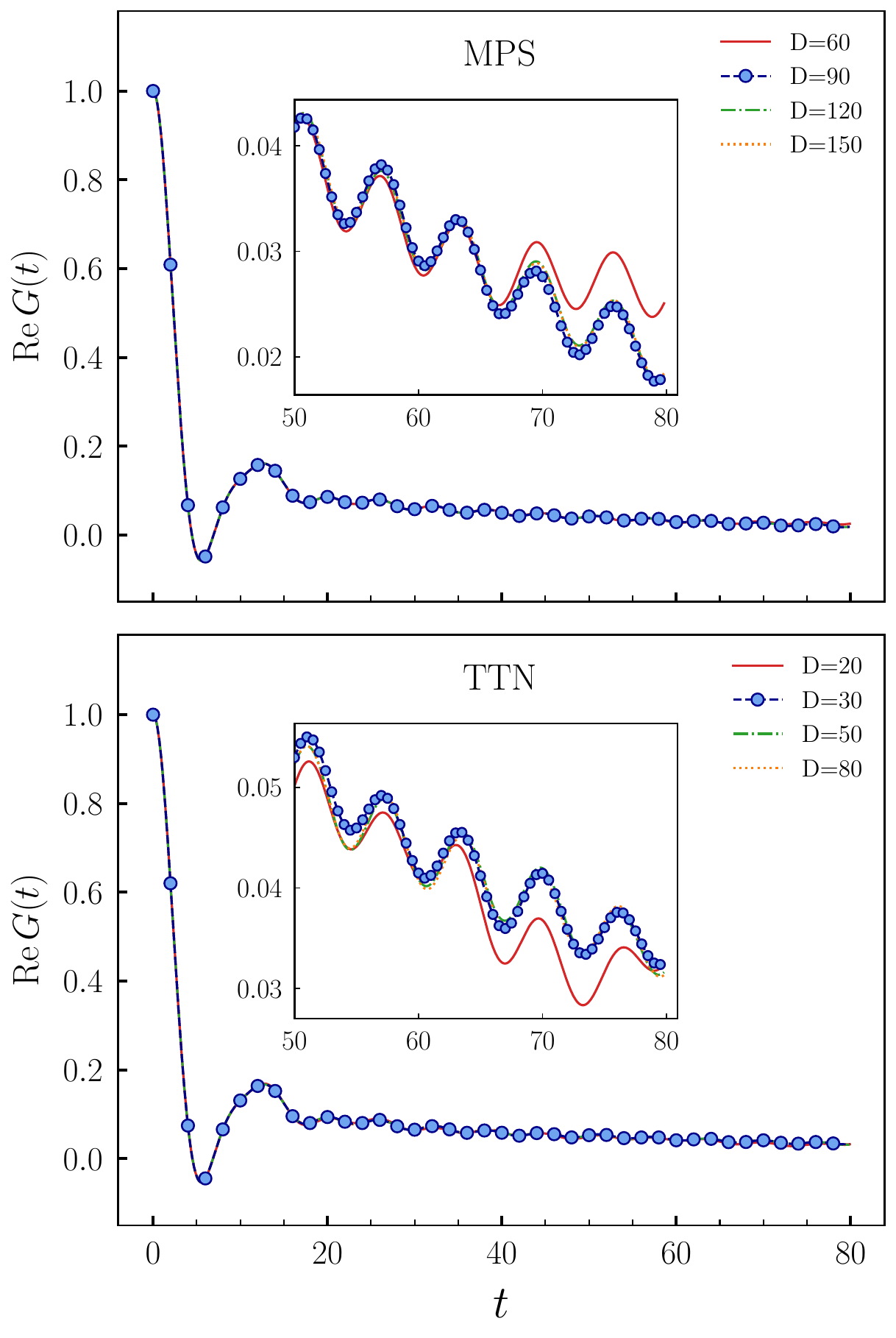}
    \caption{Time evolution of $\mathrm{Re}\,G(t)$ obtained using MPS and TTN with time step $\tau=0.1$. Systematic convergence with increasing bond dimension $D$ is observed, as indicated in the legend.}
    \label{fig:Gt_dimension_comparison}
\end{figure}

Figure~\ref{fig:Gt_dimension_comparison} illustrates the convergence of the time evolution of $\mathrm{Re}\,G(t)$ obtained using MPS and TTN as the bond dimension $D$ is increased. For the MPS, convergence of the long-time dynamics up to $t=80$ is achieved only at $D=90$. By contrast, the TTN converges at a significantly lower bond dimension, $D=30$. This difference in convergence behavior reflects the greater efficiency of the TTN representation. It motivates the choice of bond dimensions ($D=90$ for MPS and $D=30$ for TTN) used in the comparative real-time evolution analysis presented in Fig.~\ref{fig:Re_Gt} of the main text.

\section{Real-time Entanglement}
\label{sec:real_time_S_dense}

\begin{figure}[t]
    \centering
    \includegraphics[width=0.85\linewidth]{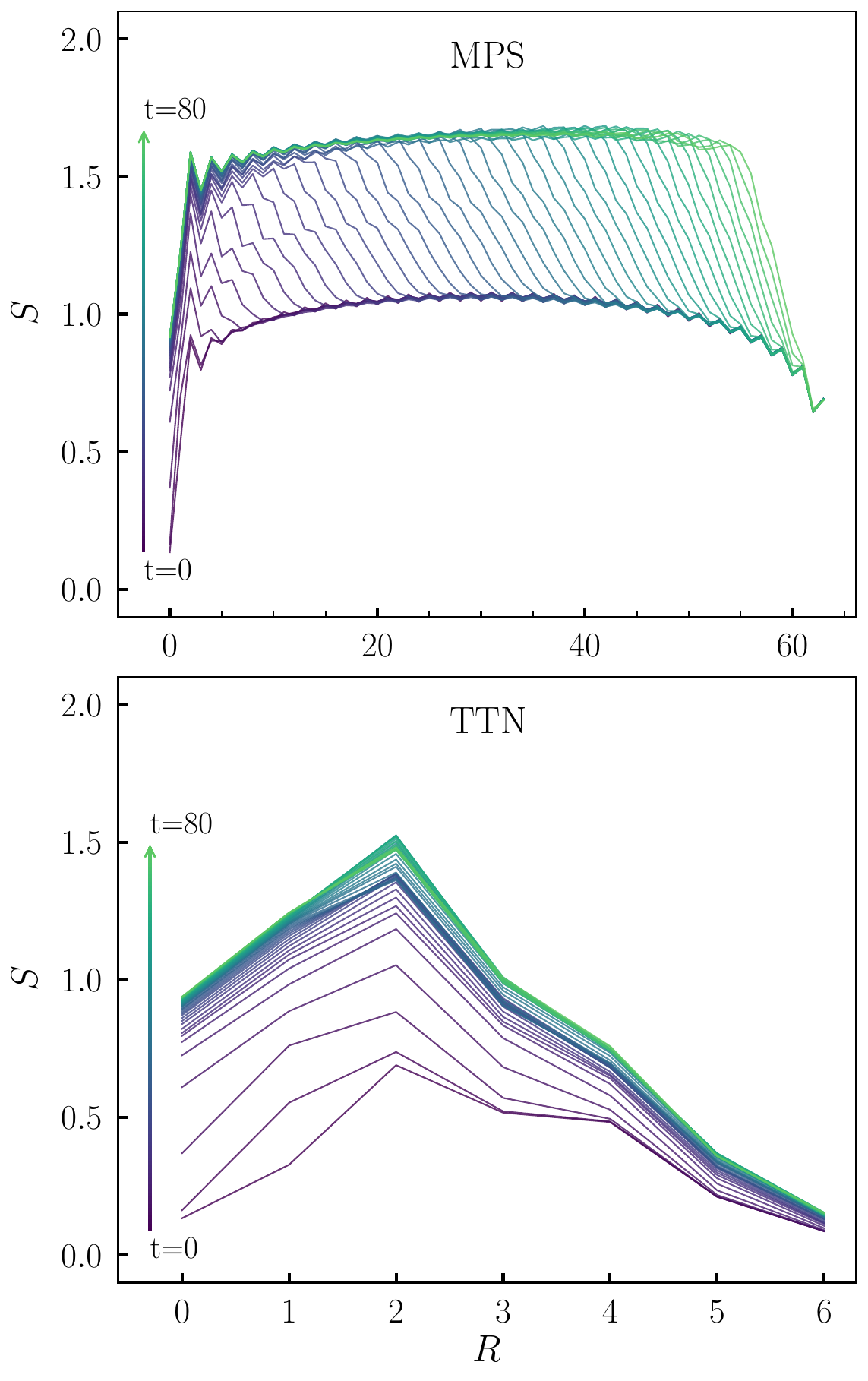}
    \caption{ 
    Time evolution of the entanglement entropy profile for MPS with $D=90$ and TTN with $D=30$ over the time interval $t=0$ to $t=80$. Time progression is indicated by the color gradient (purple to green) and arrow in the inset.
    }
    \label{fig:S_evolution_dense}
\end{figure}

Figure~\ref{fig:S_evolution_dense} shows how the entanglement entropy evolves within a dense time segment from $t=0$ to $t=80$. In both TTN and MPS, the entanglement entropy near the impurity site increases rapidly with time before approaching saturation. A characteristic feature of the MPS results [Fig.~\ref{fig:S_evolution_dense}(a)] is the emergence of a step-like entropy profile at each particular time. This step-like profile arises from the causal propagation of the local excitation along the one-dimensional chain, giving rise to a propagating entanglement front, known as the Lieb-Robinson bound~\cite{Lieb1972}. Bonds located within a time-dependent critical radius, set by the evolution time and the Lieb-Robinson velocity, develop enhanced entanglement, producing a sharp, step-like entropy profile that expands outward as time progresses.

In the TTN geometry [Fig.~\ref{fig:S_evolution_dense}(b)], the entanglement propagation respects the same causal bound but is distributed across the hierarchical structure, preventing the formation of a sharp, one-dimensional front. The entanglement at the layers centered at $R\sim 2$ approaches saturation, while the inner and outer layers continue to increase slowly with time. This persistent hierarchy enables the TTN to efficiently represent the time-evolved state.

\begin{figure}[t]
    \centering
    \includegraphics[width=0.85\linewidth]{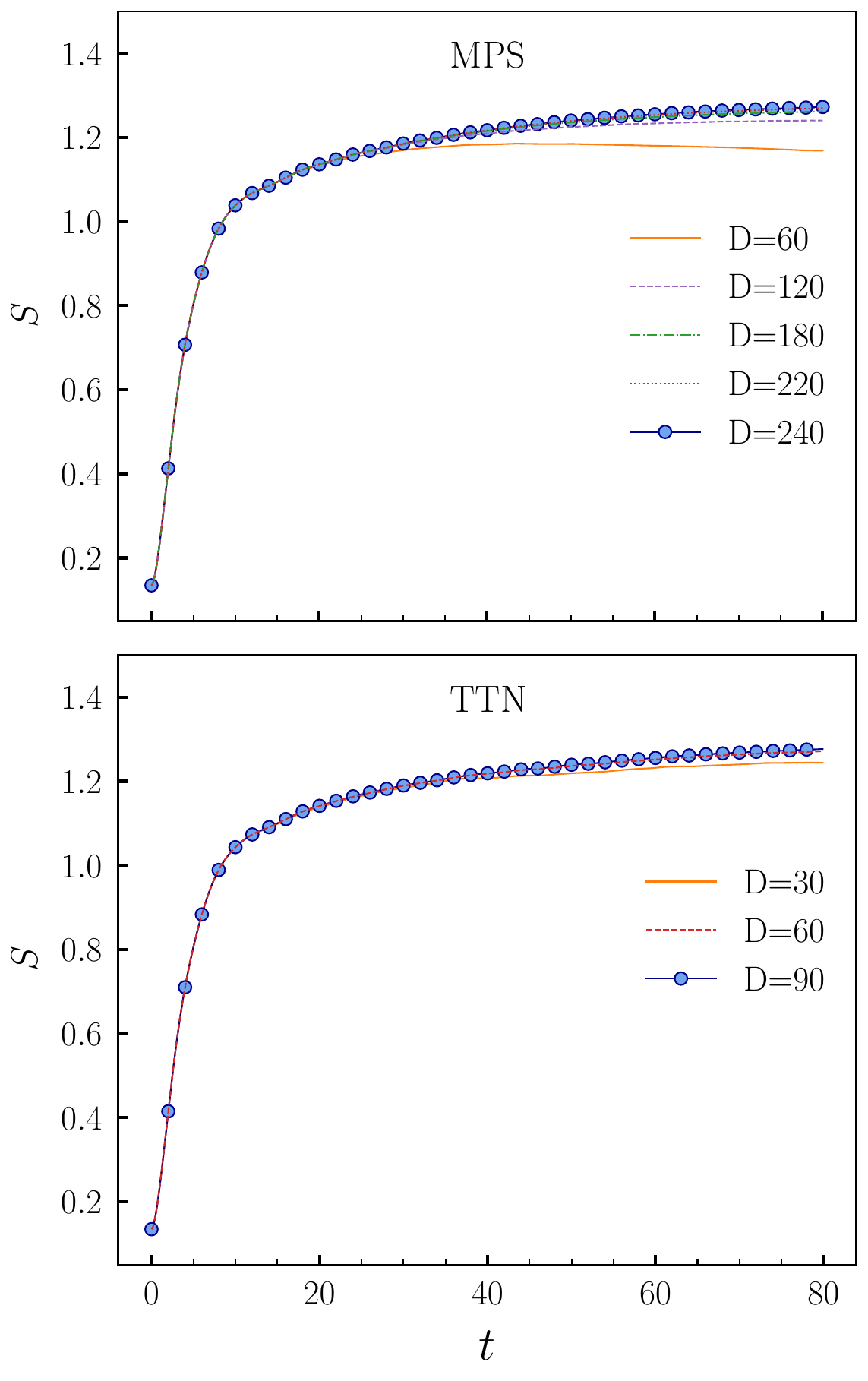}
    \caption{ 
    Time evolution of the entanglement entropy $S(t)$ at distance $R=1$ for MPS (top) and TTN (bottom) with different bond dimension $D$.
    }
    \label{fig:S_t_converge}
\end{figure}

 Figure~\ref{fig:S_t_converge} shows how the entanglement entropy $S$ at $R=1$ converges as a function of bond dimension $D$. The MPS reaches numerical convergence at about $D=240$, while the TTN achieves similar convergence at $D=90$. Therefore, we set $D=240$ for the MPS and $D=90$ for the TTN in the comparative analysis of entanglement dynamics in Fig.~\ref{fig:S_t_differentR} of the main text.

\section{Linear Prediction}
\label{Sec:LP}

Linear prediction (LP)~\cite{white2004, Barthel2009,Wolf2014_2,Ganahl2014} is a commonly used technique for extrapolating finite time series to enhance spectral resolution. It is particularly effective in the analysis of Green’s functions, where an accurate description of smooth line shapes and sharp low-energy features requires access to long-time dynamics beyond the directly computed time window. The central idea of LP is to model future data points as linear combinations of previously known values, thereby enabling controlled extrapolation of the time evolution.

 Formally, given $N_{\max}=2N_p$ uniformly sampled data points $x_i$ at times $t_i$, the LP scheme predicts values at later times $t_n$ ($n > N_{\max}$) using the relation
\begin{equation}
    \hat{x}_n = \sum_{j=1}^{N_p} a_j x_{n-j},
\end{equation}
where $\{a_j\}$ are the prediction coefficients. These coefficients are determined by minimizing the loss function
\begin{equation}
    L = \sum_{n=N_p+1}^{2N_p} \bigl| \hat{x}_n - x_n \bigr|^2,
\end{equation}
which measures the deviation between the predicted and actual data over the training window. By minimizing $L$ with respect to $a_k^*$, one obtains the normal equations
\begin{equation}
    R \vec{a} = \vec{r},
\end{equation}
where the correlation matrix $R$ and the cross-correlation vector $\vec{r}$ are defined as
\begin{align}
    R_{ij} &= \sum_{n=N_p+1}^{2N_p} x_{n-i}^* x_{n-j}, \\
    r_i &= \sum_{n=N_p+1}^{2N_p} x_{n-i}^* x_n .
\end{align}

The coefficient vector $\vec{a}$ is obtained by solving this linear system via pseudoinversion, $\vec{a}=R^{-1}\vec{r}$, with a cutoff parameter $\delta$ introduced to regularize small singular values and ensure numerical stability. This procedure enables reliable extrapolation while preserving the dominant spectral characteristics of the original time series.

The LP dynamics can be expressed in terms of a state-space representation,
\begin{equation}
    s_{k+1} = M s_k ,
\end{equation}
where the state vector is defined as $s_k = (x_k, x_{k-1}, \dots, x_{k-N_p+1})^{\mathrm{T}}$ and the transition matrix $M$ takes the form
\begin{equation}
  M =
  \begin{pmatrix}
    a_1 & a_2 & a_3 & \dots & a_{N_p} \\
    1   & 0   & 0   & \dots & 0 \\
    0   & 1   & 0   & \dots & 0 \\
    \vdots & \ddots & \ddots & \ddots & \vdots \\
    0   & 0   & \dots & 1 & 0
  \end{pmatrix}.
\end{equation}

If $M$ possesses eigenvalues $\lambda$ with magnitude $|\lambda| > 1$, the corresponding modes grow exponentially with increasing $k$, leading to unphysical divergence in the predicted time evolution. To suppress such instabilities, the transition matrix $M$ is renormalized by eliminating or damping eigenmodes with $|\lambda|>1$, for example by setting these eigenvalues to zero~\cite{Ganahl2014, Ganahl2015}. This procedure preserves the physically relevant decaying modes of the Green’s function while removing spurious growing contributions, resulting in stable and physically meaningful long-time extrapolations.

\section{Trotter error in the free-fermion model}
\label{sec:appendix_trotter}

To further verify the geometric advantage of the Cayley-tree mapping in suppressing Trotter errors, we consider a two-dimensional non-interacting fermion system on an $8\times 8$ square lattice with nearest-neighbor hopping and periodic boundary conditions. By treating one site as the impurity, we map the remaining bath degrees of freedom onto both a 1D chain and a Cayley tree.

\begin{figure}[b]
    \centering
    \includegraphics[width=0.9\linewidth]{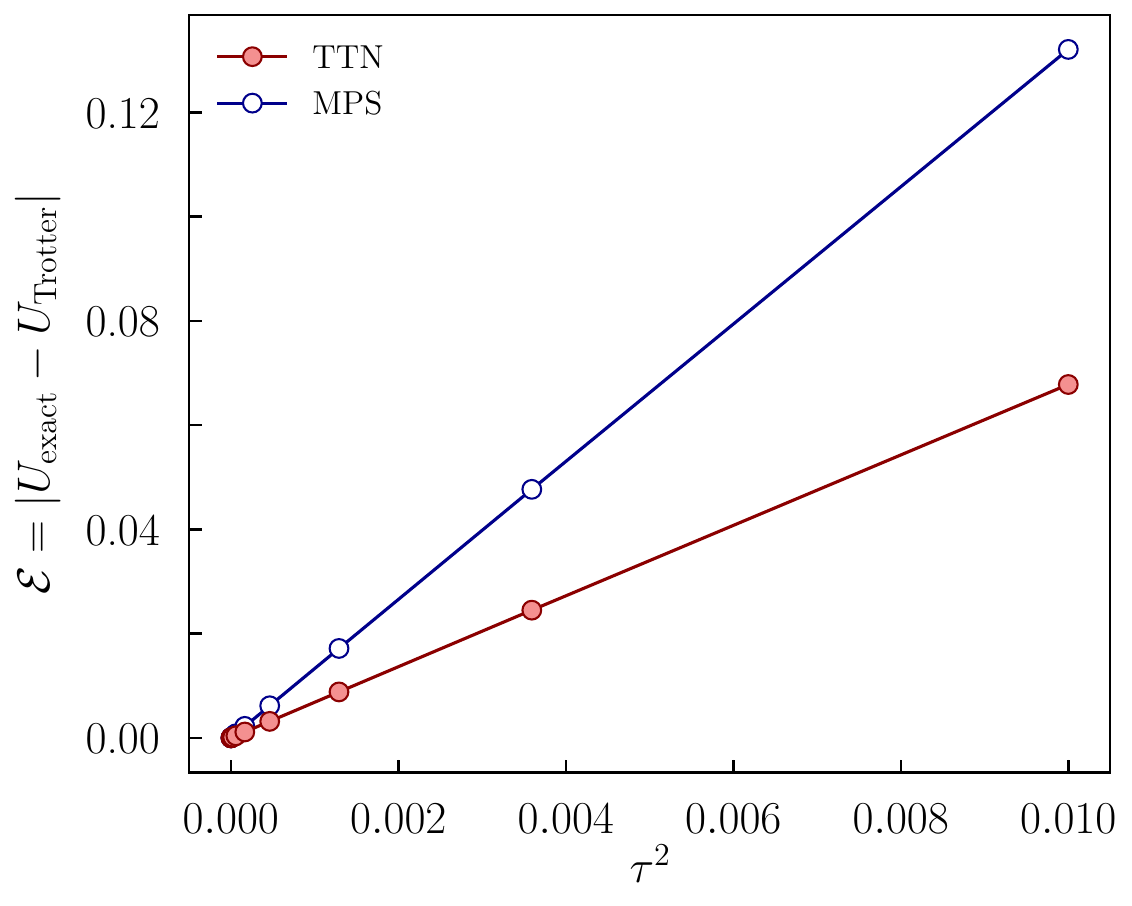}
\caption{
Trotter error $\mathcal{E}=|U_{\mathrm{exact}}-U_{\mathrm{Trotter}}|$ versus $\tau^{2}$ for the chain mapping and the Cayley-tree mapping of the free-fermion model on a periodic $8\times8$ square lattice. $U=\exp(i\tau H)$ is the time-evolution operator.}
    \label{fig:trotter_free}
\end{figure}

Because the system is noninteracting, the time-evolution operator $U(\tau) = e^{iH\tau}$ can be computed exactly. This allows us to evaluate the Trotter error without contamination from bond-dimension truncation.  As shown in Fig.~\ref{fig:trotter_free}, both mappings exhibit the expected $\mathcal{O}(\tau^2)$ scaling of the Trotter error. However, the Cayley-tree mapping  yields a significantly smaller prefactor than the chain mapping. A quantitative fit gives slopes of 6.8 and 13.3 for the Cayley-tree and chain mappings, respectively. This clearly demonstrates that the rapid geometric decay of the hopping amplitudes in the tree mapping more effectively suppresses the accumulation of Trotter errors.

\bibliography{ref}

\end{document}